%% file: ACC_ArXiv.tex
\documentclass[preprint,1p,11pt]{ISAS_IR}

\usepackage{ISASPackages/ISASMatheMacros}

\usepackage[english]{babel}
\usepackage[latin1]{inputenc}
\usepackage{calc}
\usepackage{amsmath}
\usepackage{amssymb}
\usepackage{microtype}
\usepackage{lmodern}
\usepackage{theorem}
\theoremheaderfont{\bfseries}
\theoremstyle{plain}
\usepackage{tikz}
\usepackage{pgfplots}
\usepackage{booktabs}
\usepackage{subfigure}
\usepackage{colortbl}
\usetikzlibrary{positioning}
\usetikzlibrary{shapes,shadows}
\tikzstyle{block} = [draw, fill=blue!10, rectangle, 
    minimum height=2em, minimum width=2em, text width= 4.5em, text centered, drop shadow, rounded corners]
\tikzstyle{entry} = [draw, rectangle, 
    minimum height=2em, minimum width=2em, text width=3em, text centered]
\tikzstyle{pinstyle} = [pin edge={to-,thin,black}]
\newtheorem{theorem}{Theorem}
\newtheorem{remark}{Remark}

\newtheorem{lemma}{Lemma}

\newcommand{\tr}{^{\rm T}}

\newcommand{\no}{\nonumber}

\input{Defs.h} 

\begin{document}

\begin{frontmatter}

\title{\LARGE \bf Optimal Sequence-Based LQG Control over TCP-like Networks Subject to Random Transmission Delays \\ and Packet Losses}

\author[isas]{J\"org Fischer}
\ead{joerg.fischer@kit.edu}

\author[isas]{Achim Hekler}
\ead{achim.hekler@kit.edu}

\author[isas]{Maxim Dolgov}
\ead[isas]{maxim.dolgov@student.kit.edu}

\author[isas]{Uwe D. Hanebeck}
\ead[isas]{uwe.hanebeck@iee.org}

\address[isas]{Intelligent Sensor-Actuator-Systems Laboratory (ISAS)\\
Institute for Anthropomatics\\
Karlsruhe Institute of Technology (KIT), Germany}

\begin{abstract}
    This paper addresses the problem of sequence-based controller design for Networked Control Systems (NCS), where control inputs and measurements are transmitted over TCP-like network connections that are subject to stochastic packet losses and time-varying packet delays. 
    At every time step, the controller sends a sequence of predicted control inputs to the actuator in addition to the current control input. 
    In this sequence-based setup, we derive an optimal solution to the Linear Quadratic Gaussian (LQG) control problem
    and prove that the separation principle holds. Simulations demonstrate the improved performance of this optimal controller compared to other sequence-based approaches.
\end{abstract}

\end{frontmatter}

%
%
 \section{INTRODUCTION}
 \label{sec:introduction}
 \input{10_introduction}


 \section{SYSTEM SETUP \& PROBLEM FORMULATION}
 \label{sec:setup}
 \input{20_setup}
 
 \section{DERIVATION OF THE OPTIMAL CONTROLLER}
 \label{sec:controller}
 \input{30_controller}

 \section{SIMULATIONS}
 \label{sec:simulations}
\input{50_simulations}

 \section{CONCLUSIONS}
 \label{sec:conclusions}
 \input{60_conclusions}


\bibliographystyle{IEEEtran}
\bibliography{literature}

\end{document}

%% file: 10_introduction.tex
In Networked Control Systems (NCS), components of a control loop are connected by one or more digital data networks. The applied data networks can be distinguished into two groups. There are real-time capable fieldbuses, such as Interbus, PROFIBUS, or Ether-CAT, which guarantee reliable data transmissions with deterministic latency on the one hand and general-purpose networks such as Ethernet-(TCP/UDP)/IP, WPAN (IEEE 802.15) or WLAN (IEEE 802.11) that have a stochastic transmission characteristic on the other hand. While fieldbuses have been the standard in industrial control systems for more than two decades, there is a trend towards applying general-purpose networks within the control loop for several reasons. Networks like Ethernet-TCP/IP are not only cheaper than the commercial fieldbuses but are also based on a wide-spread, nonproprietary, and standardized communication technology \cite{antsaklis2007special}. Furthermore, wireless networks are much more flexible than, e.g., the wired ring topology of Interbus, and allow for applications not realizable with fieldbuses such as mobile sensor networks \cite{Seiler05} and freely moving objects \cite{Ogren04}.

However, general-purpose networks can be subject to highly time-varying transmission delays and to immense data losses.
These effects can strongly degrade the performance of a system and even destabilize the control loop \cite{Zhang01, Bemporad10, Heemels10}. 
Therefore, new control methods have been developed that take the stochastic network effects explicitly into account~\cite{Hespanha07}. 

\begin{figure}
	\centering
		\includegraphics[width=0.7\textwidth]{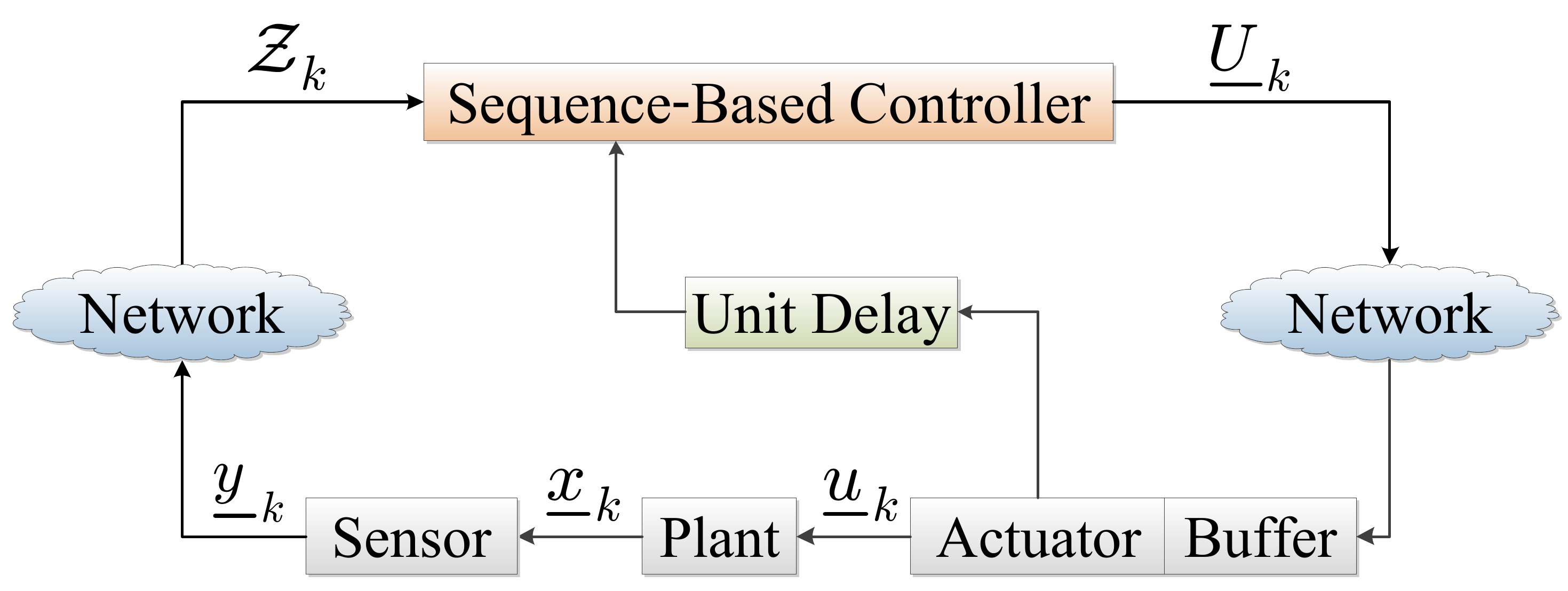}
	\caption{Considered setup: The sequence-based controller generates control input sequences $\vec{U}_k$ that are 
transmitted to the actuator via a TCP-like network connection that provides acknowledgements for successfully transmitted data packets. The actuator holds the most recent control input sequence in a buffer and applies the time-corresponding entry $\vec{u}_k$ of the sequence to the plant. The state $\vec{x}_k$ of the plant is measured ($\vec{y}_k$) and sent via a network to the controller, which receives no, one, or even more than one measurement ($\mathcal{Z}_k$).}
	\label{fig:system_overview}
\end{figure}

Our approach belongs to a class of methods called sequence-based control, which is, depending on the author, also referred to as networked predictive control, packet-based control or receding horizon networked control \cite{Bemporad98,Gruene09,Quevedo11, Tang07, Liu07, gupta2006receding}. The main idea of this approach is that the controller sends data packets over the network to the actuator that contain not only the current control input, but also predicted control inputs for future time instants. The predicted future control inputs are stored in a buffer attached to the actuator so they can be applied in cases future data packets are delayed or lost. An assumption made by these methods is that the additional data sent over the network does not degrade the quality of the connection. For packet-based networks such as Ethernet-TCP/IP, this assumption usually holds since the process data needed for control applications is normally much smaller than the size of a data packet. 

Depending on the considered system, different approaches for design of sequence-based controllers have been proposed. One class of these approaches, as for example used in \cite{Liu07, Liu06, HeklerVI12, HeklerCDC12}, is based on a nominal feedback-controller that is designed for the nominal system, where the networks are replaced by transparent connections. Using the nominal controller, future control inputs are predicted and a sequence-based controller is synthesized. For linear systems, this approach can provide stability of the system, however, in the presence of packet losses, the resulting controller is in general not optimal even if the nominal controller is the result of an optimization method, such as LQG, ${\rm H_2}$, or ${\rm H_{\infty}}$.

Another line of sequence-based approaches evolves from Model Predictive Control (MPC) theory (\cite{Gruene09, Quevedo11, Quevedo07}). This is an intuitive connection since MPC-controllers already obtain an optimized sequence of predicted control inputs over a finite horizon. In standard MPC, only the first control input of this sequence is applied to the system and the rest is discarded. If instead the whole sequence is lumped into a data packet and sent to the actuator, a sequence-based controller is derived.
However, like standard MPC, this approach is suboptimal since the real closed-loop optimization problem is approximated by the much easier open-loop optimization problem, which is suitable for systems that are far too complex to derive optimal solutions.

In \cite{gupta2006receding}, it has been shown, based on \cite{gupta2006stochastic} and \cite{sinopoli2004kalman}, that optimal sequence-based controllers in the context of Linear Quadratic Gaussian (LQG) control can be derived for networks that use a so called TCP-like protocol. The term TCP-like\footnote{A TCP-like network connection is only an approximation of a realistic Ethernet-TCP/IP network since it is assumed that the acknowledgements are not subject to time delays. However, the analysis of NCS with TCP-like connections gives insights into the more complex problem of controlling systems over real TCP connections or even over networks that do not provide any acknowledgments, such as networks using a UDP protocol. In particular, the TCP-like case constitutes an upper performance bound for NCS with realistic TCP or UDP network connections.} characterizes a data network that provides instantaneous acknowledgments for successfully transmitted data packets \cite{Schenato07}.
Yet, the approach in \cite{gupta2006receding} neglects the possibility of time delays by assuming that a data packet is either dropped by the network or transmitted immediately. In \cite{moayedi2011lqg}, the approach was extended so that time delays could be incorporated into the sequence-based control design. However, the derived controller is not optimal since the authors artificially limit the information available to the controller for calculating control input sequences.

In this paper, we present the optimal solution to the problem addressed in \cite{gupta2006receding} and \cite{moayedi2011lqg} also in the presence of time-varying transmission delays. In detail, we derive an optimal solution for the sequence-based LQG control problem for NCS with TCP-like network connections subject to stochastic packet losses and time-varying packet delays. The considered setup is depicted in fig.~\ref{fig:system_overview}. The plant is assumed to be partially observable, discrete-time and linear and perturbed by additive process and measurement noise. 
The optimal control law is derived by first using state augmentation to formulate the original networked system as a nonnetworked Markovian Jump Linear System (MJLS). Then, stochastic dynamic programming is applied on the MJLS. In the derivation, we prove that the separation principle holds also in the sequence-based setup, what was assumed in former work but not formally proved.

\subsection{Notation}
Throughout the paper, random variables such as $\rv{a}$ are written in bold face letters, whereas deterministic quantities $a$ are in normal lettering.
Furthermore, the notation $\rv{a} \sim f(a)$ refers to a random variable $\rv{a}$ with probability density function $f(a)$.
A vector-valued quantity $\vec{a}$ is indicated by underlining the corresponding identifier and matrices are always referred to with bold face capital letters, e.g., $\mat{A}$.
The notation $a_k$ refers to the quantity $a$ at time step $k$. 
%
%
For the set $\left\{ \vec{x}_a, \vec{x}_{a+1}, \dots, \vec{x}_{b}\right\}$, we use the abbreviated notation $\vec{x}_{a:b}$.
The expectation operator is denoted by $\E\{\cdot\}$ and the Moore-Penrose pseudoinverse of a matrix $\mat{A}$ by $\mat{A}^{\dagger}$.
The set of all natural numbers including zero is indicated with $\mathbb{N}_0$ and we use the symbol $\mathbb{N}_{>0}$ for $\mathbb{N}_0 \backslash \{ 0 \}$.
%
\subsection{Outline}
The remainder of the paper is organized as follows.
In the next section, the considered setup is introduced and the optimal control problem is formulated. 
The optimal control law is derived in sec.~\ref{sec:controller} and compared with the approaches from \cite{gupta2006receding} and \cite{moayedi2011lqg} in a simulation of a double integrator system in sec.~\mbox{\ref{sec:simulations}}. 
A summary and an outlook on future work concludes the paper.

%% file: 20_setup.tex
We consider the system setup depicted in fig.~\ref{fig:system_overview}. The partially observable plant and the sensor evolve according to the stochastic, discrete-time, linear equations
\begin{align}
  \rvec{x}_{k+1} &= \mat{A} \rvec{x}_k + \mat{B} \vec{u}_k + \rvec{w}_k \ , \label{eq:sysX} \\
	\vec{y}_{k}   &= \mat{C} \rvec{x}_k + \rvec{v}_k 										 \ , \label{eq:sysY}
\end{align}
where $\rvec{x}_k \in \mathbb{R}^n$ denotes the plant state at time step $k$, the vector $\vec{u}_k \in \mathbb{R}^m$ the control input applied by the actuator, and $\vec{y}_k \in \mathbb{R}^q$ the measured output. The matrices \mbox{$\mat{A} \in \mathbb{R}^{n \times n}$}, \mbox{$\mat{B} \in \mathbb{R}^{n \times m}$}, and \mbox{$\mat{C} \in \mathbb{R}^{q \times n}$} are known and we assume that $\left(\mat{A}, \mat{B}\right)$ is controllable and $\left(\mat{A}, \mat{C}\right)$ is observable. The terms $\rvec{w}_k \in \mathbb{R}^n$ and $\rvec{v}_k \in \mathbb{R}^q$ represent mutually independent, stationary, zero-mean, discrete-time, white noise processes with Gaussian probability distribution
that are independent of network-induced effects. The initial condition of the state $\rvec{x}_0$ is independent of the other random variables and has a Gaussian distribution. We define 
$$\bar{\vec{x}}_0 = E\left\{ \rvec{x}_0 \right\} \ \ \text{and} \ \ \overline{\mat{P}}_0 = \E \left\{ \left(\rvec{x}_0 - \bar{\vec{x}}_0 \right) \left(\rvec{x}_0 - \bar{\vec{x}}_0 \right)\tr\right\} .$$
%
The data connections between controller and actuator (CA-link) and between sensor and controller (SC-link) are provided by data networks that are subject to time-varying delays and stochastic packet losses. By interpreting lost transmissions as transmissions with infinite time delay, we unify the description of both effects by only considering time-varying but possibly unbounded time delays. The time delays are described by random processes $\rv{\tau}^{C\!A}_k, \rv{\tau}^{SC}_k \in \mathbb{N}_0$ that specify how many time steps a data packet will be delayed if sent at time step $k$. Throughout the paper, we assume that $\rv{\tau}^{C\!A}_k$ and $\rv{\tau}^{SC}_k$ are white stationary processes and that their discrete probability density functions $f^{SC}(\tau^{SC}_k)$ and $f^{C\!A}(\tau^{C\!A}_k)$ are known.
Additionally, it is assumed that the components of the control loop are time-triggered, time-synchronized and have identical cycle times. Furthermore, the employed network is capable of transmitting large time stamped data packets and uses a TCP-like protocol, i.e., acknowledgements are provided within the same time step when a packet was successfully transmitted.

Every time step, the measurement $\vec{y}_k$ is sent to the controller. Due to transmission delays and packet losses, it is possible that the controller receives no, one, or even more than one measurement. The set of received measurements at time step $k \in \mathbb{N}_{>0}$ is defined as the set $\mathcal{Z}_k$ according to
\begin{equation}
	\mathcal{Z}_k = \left\{\vec{y}_m : m \in \left\{0, 1, \cdots, k\right\}, m + \tau_k^{SC} = k \right\} \label{eq:zk} \ .
\end{equation}

After processing $\mathcal{Z}_k$, the sequence-based controller generates a control input sequence $\vec{U}_k$ that is sent over the network to the actuator. Entries of that sequence are denoted by $\vec{u}_{k+m|k}$ with $m \in \{0, 1, ..., N\}$ and $N \in \mathbb{N}_0$. The index specifies that the control input is intended to be applied at time step $k+m$ and was generated at time step $k$. This way, a sequence of length $N + 1$ generated at time step $k$ is described by
\begin{equation}
	\vec{U}_k = \begin{bmatrix}\vec{u}_{k|k}\tr & \vec{u}_{k+1|k}\tr & \dots & \vec{u}_{k+N|k}\tr \end{bmatrix}\tr \ . \label{eq:sequenceEntries}
\end{equation}
Attached to the actuator is a buffer, in which the actuator stores the sequence with the most recent information among all received sequences, i.e., the sequence that was generated last (according to the time stamps). If a sequence arrives out of order, the actuator does not change the content of the buffer and the received sequence is discarded. After updating the buffer, the actuator applies the control input of the buffered sequence that corresponds to the current time step. Since we do not assume that the time delays are bounded, it may happen that the buffer runs empty. In this case the controller applies a time-invariant default control input $\vec{u}^d$. Furthermore, the actuator initializes the buffer at time step $k=0$ with a sequence of default control inputs. The described actuator procedure can formally be summarized by
\begin{align}
	\vec{u}_k &= \vec{u}_{k|k-\rv{\theta}_k} \ , \label{eq:uk} \\
	\rv{\theta}_k &= \min \left( \left\{ n \in \mathbb{N}_0 : m + \rv{\tau}_m^{CA} = k - n, m \in \mathbb{N}_0 \right\}   \cup \left\{ N + 1 \right\} \right) \ ,  \label{eq:thetak}\\
  					& \hspace{-0.3cm} \vec{u}_{k|k-N-1} =  \vec{u}^d \ \label{eq:ud}.
\end{align}
%
%
\begin{remark}
\label{rem:theta}The random variable $\rv{\theta}_k$ can be interpreted as the \textit{age} of the sequence buffered in the actuator, i.e., the difference between time step of generation and actual time step. If no appropriate control input is buffered in the actuator, $\rv{\theta}_k$ is set to $N+1$ and the default control input $\vec{u}^d$ is applied according to \eqref{eq:ud}.
\end{remark}
As we consider a TCP-like protocol, the controller can always infer which control input has been applied to the plant. 
Therefore, at time step $k$, the controller has access to the past realizations of $\rv{\theta}_k$. Based on the previous passages, we can summarize the information available to the controller at time step $k$ by the information set $\mathcal{I}_k$ according to
\begin{equation}
	\mathcal{I}_k = \left\{ \bar{\vec{x}}_0, \overline{\mat{P}}_0, \mathcal{Z}_{1:k}, \vec{U}_{0:k-1}, \theta_{0:k-1} \right\} \ \label{eq:Ik}.
\end{equation}
At every time step $k$, the controller maps the available information to a control input sequence $\vec{U}_k$. Denoting the mapping function at time step $k$ with $\mu_k$, the control law is given by the set of all $\mu_k$ with $k \in \left\{0, 1, \dots, K-1 \right\}$, where $K$ is the terminal time. A control law is called admissible if
\begin{equation}
	\vec{U}_k = \mu_k(\mathcal{I}_k) \label{eq:mu} \ , \  \ \forall \ k \in \left\{ 0, 1, \dots , K-1 \right\}
\end{equation}
holds. In this paper, we are interested in finding an admissible control law that minimizes the cumulated linear quadratic cost function
\begin{align}
  C_0^K = {\E} \left\{ C_K + \displaystyle\sum_{k=0}^{K-1} C_k \Big{|} \vec{U}_{0:K-1}, \bar{x}_0, \overline{\mat{P}}_0\right\}  \ \label{eq:costFunction},
\end{align}
with stage cost
\begin{align}
	C_K &= \rvec{x}_K\tr \mat{Q}_K \rvec{x}_K  \ , \label{eq:stageCostN}\\
	C_k &= \rvec{x}_k\tr \mat{Q}_k \rvec{x}_k + \vec{u}_k\tr \mat{R}_k \vec{u}_k \label{eq:stageCostk}\ ,
\end{align}
where $K \in \mathbb{N}_{>0}$ is the terminal time step, $\mat{Q}_k$ is positive semidefinite, and $\mat{R}_k$ is positive definite.

Summarizing the optimal control problem, we seek to find an admissible control law according to \eqref{eq:mu} that minimizes the cost \eqref{eq:costFunction} - \eqref{eq:stageCostk} 
subject to the system dynamics \eqref{eq:sysX}, the measurement equations \eqref{eq:sysY}, \eqref{eq:zk}, and the actuator \mbox{logic \eqref{eq:uk} - \eqref{eq:ud}}.

%% file: 30_controller.tex
In order to derive the optimal controller, we model the system as a MJLS in sec.~\ref{sub:model}. Based on this model, we derive the optimal controller via stochastic dynamic programming in sec.~\ref{sub:sdp}.
\subsection{System Modeling}
\label{sub:model}
According to \eqref{eq:uk} and \eqref{eq:ud}, the control input applied by the actuator at time step $k$ is given by
\begin{equation}
	\vec{u}_k = \vec{u}_{k|k-\rv{\theta}_k} \ , \ \ \ \ \  \vec{u}_{k|k-N-1} =  \vec{u}^d \ ,
\end{equation}
where $\rv{\theta}_k$ is described by \eqref{eq:thetak}. 
As mentioned in remark \ref{rem:theta}, the random variable $\rv{\theta}_k$ can be interpreted as the \textit{age} of the sequence buffered in the actuator. Furthermore, it has been shown in \cite{HeklerVI12} and \cite{Fischer12} that $\rv{\theta}_k$ can be described as state of a Markov chain with transition matrix $\mat{T}$ according to
\begin{align}
	\mat{T} =  \label{eq:P}
	\begin{bmatrix}
		p_{00} & p_{01} & 0      & 0  & \cdots & 0\\
		p_{10} & p_{11} & p_{12} & 0  & \cdots & 0\\
		p_{20} & p_{21} & p_{22} & p_{23} & \cdots & 0\\
		\vdots & \vdots & \vdots & \vdots & \ddots & \vdots\\
		\vdots & \vdots & \vdots & \vdots & \vdots & p_{(r-1)(r)}\\
		p_{r0} & p_{r1} & p_{r2} & p_{r3} & \cdots & p_{rr}
	\end{bmatrix}\ , 
\end{align}
with
\begin{align}
	p_{ij} ={\rm Prob}\left[\rv{\theta}_{k+1} = j|\rv{\theta}_k = i \right], \nonumber \ \  \ r=N+1 \ .
\end{align}
The elements of $\mat{T}$ in the upper right triangle are zero as $\rv{\theta}_k$ can only increase by one per time step. The remaining entries can be calculated by
\begin{align*}
	p_{(i-1)(i)} &= 1 - \sum\limits_{s=0}^{i-1} q_s \ \ \ \ \ \text{for} \ i \in \left[1, 2, \cdots, N + 1 \right] \ , \\
  p_{ij} &= q_j \ \ \ \ \text{for} \ j \leq i \ , \text{where} \ i, j \in \left[0, 1, \cdots, N + 1 \right] ,
\end{align*}
where $q_i$ is the probability that a sequence is delayed for $i \in \mathbb{N}_0$ time steps. The $q_i$'s can directly be derived from $f^{C\!A}(\tau_k^{C\!A})$. For a detailed derivation of the transition matrix we refer to \cite{HeklerVI12}.

\begin{figure}
	\centering
		\includegraphics[width=0.8\textwidth]{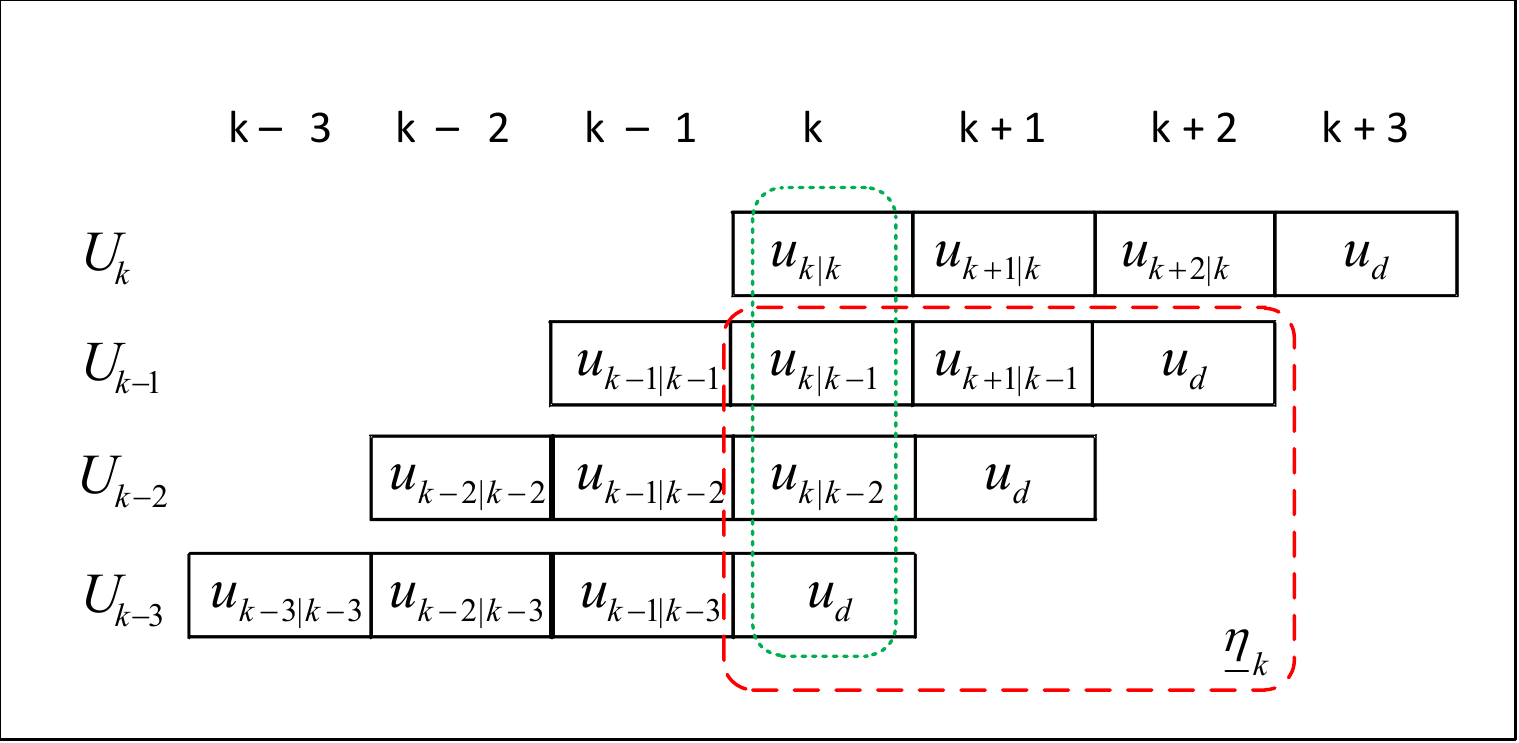}
	\caption{Representation of control input sequences $\vec{U}_{k-3}, \hdots, \vec{U}_k$, whereas control inputs corresponding to the same time
step are vertically aligned. The default control $\vec{u}^d$ is added to the end of every sequence. The control inputs that could be applied by the actuator
at time step $k$ are marked by the green doted rectangle and the control inputs that are part of $\vec{\eta}_k$ are marked by the red dashed rectangle.}
	\label{fig:packets}
\end{figure}

With $\rv{\theta}_k$, we can describe which sequence is buffered at time step $k$. To summarize all control inputs of sequences that could be buffered, we introduce the vector
\begin{equation}
	\begin{aligned}
		\vec{\eta}_k &=
		\begin{pmatrix}
			[\vec{u}_{k|k-1}\tr \ \ \vec{u}_{k+1|k-1}\tr \ \ \cdots \ \ \vec{u}_{k+N-1|k-1}\tr ]\tr \\
			[\vec{u}_{k|k-2}\tr \ \ \vec{u}_{k+1|k-2}\tr \ \ \cdots \ \ \vec{u}_{k+N-2|k-2}\tr]\tr \\
			\ \ \ \ \vdots\\
			\ \ \ \ \ \ [\vec{u}_{k|k-N+1}\tr \ \ \vec{u}_{k+1|k-N+1}\tr]\tr \\
			\ \ \ \ \ \vec{u}_{k|k-N} \\
			\ \ \ \ \ \vec{u}^d
		\end{pmatrix} \ ,
	\end{aligned}
	\label{eq:eta} 
\end{equation}
with $\vec{\eta}_k \in \mathbb{R}^d$ and $d = m + m \cdot \sum_{i=1}^{N} i$. The vector $\vec{\eta}_k$
contains the default control input $u^d$ and all control inputs of the former sent sequences $\vec{U}_{k-1}, \cdots, \vec{U}_{k-N}$ that still could be applied by the actuator either in the current time step or in the future. This is illustrated in fig.~\ref{fig:packets}, where the relevant control input sequences are depicted for the case of $N = 2$. Therefore, the control input applied by the actuator at time step $k$ is either part of $\vec{\eta}_k$ or of $U_k$.

Combining $\vec{\eta}_k$ and $\rv{\theta}_k$, the following state space model of network and actuator can be derived
\begin{eqnarray}
	\vec{\eta}_{k+1} & = & \mat{F} \vec{\eta}_k + \mat{G} \vec{U}_k \label{eq:netwAndAct1} \ , \\
  {\rvec{u}}_k  & = & \mat{H}_{k} \vec{\eta}_k + \mat{J}_{k} \vec{U}_k \label{eq:netwAndAct2} \ ,
\end{eqnarray}
with
\begin{equation*}
	\begin{aligned}
		\mat{F} &= 
		\begin{bmatrix}
			\mat{0} & \mat{0} & \mat{0} & \mat{0} &\cdots & \mat{0} \\
			\mat{0} & \mat{I} & \mat{0} & \mat{0} &\cdots & \mat{0} \\
			\mat{0} & \mat{0} & \mat{0} & \mat{I} &\cdots & \mat{0} \\
			\vdots & \vdots & \vdots & \vdots & \ddots & \vdots \\
			\mat{0} & \mat{0} & \mat{0} & \mat{0} &\cdots & \mat{I} \\
		\end{bmatrix} \ ,
	\end{aligned}
	\begin{aligned}
	  & \mat{J}_{k} = 
		\begin{bmatrix}
  		\delta_{(\rv{\theta}_k,0)}\,\mat{I} & \mat{0}
		\end{bmatrix} \ , \ \
		\mat{G} =
		\begin{bmatrix}
			\mat{0} & \mat{I}\\
			\mat{0} & \mat{0}\\
		\end{bmatrix} \ , \\
	  & \mat{H}_{k} =
  	\begin{bmatrix}
  		\delta_{(\rv{\theta}_k,1)}\,\mat{I} &
  		\mat{0}&  
			\delta_{(\rv{\theta}_k,2)}\,\mat{I} &
			\mat{0}&
			\cdots &
			\delta_{(\rv{\theta}_k,N)}\,\mat{I}
  	\end{bmatrix} \ .  
	\end{aligned}
\end{equation*}
Thereby, the terms \mat{0} denote matrices with all elements equal to zero and \mat{I} the identity matrix, each of appropriate dimension. The expression $\delta_{(\rv{\theta}_k, i)}$ is the Kronecker delta, which is defined as
\begin{eqnarray*}
	\delta_{(\rv{\theta}_k, i)} = 
	\left\{ \begin{array}{rcl}
		1 & \mbox{if} & \rv{\theta}_k = i\\
		0 & \mbox{if} & \rv{\theta}_k \neq i
	\end{array}\right. \ .
\end{eqnarray*}
Defining the augmented state 
\begin{equation}
	\rvec{\xi}_k = \begin{bmatrix} \rvec{x}_k\tr & \vec{\eta}_k\tr \end{bmatrix}\tr \label{eq:augState}
\end{equation}
and combining (\ref{eq:sysX}), (\ref{eq:netwAndAct1}), and (\ref{eq:netwAndAct2}), it holds that
\begin{align}
\rvec{\xi}_{k+1} & = 
	\begin{bmatrix}
		\mat{A} & \mat{B}\cdot \mat{H}_{k}\\
		\mat{0}  & \mat{F}
	\end{bmatrix}
	\rvec{\xi}_k +
	\begin{bmatrix}
		\mat{B}\cdot \mat{J}_{k}\\
		\mat{G}
	\end{bmatrix}
	\vec{U}_k + 
	\begin{pmatrix}
		\rv{w}_k \\
		\mat{0}
	\end{pmatrix} 
	\label{eq:xi1} \\
	& = {\widetilde{\mat{A}}}_{k} \rvec{\xi}_k + {\widetilde{\mat{B}}}_{k} \vec{U}_k + \rvec{\widetilde{w}}_k \ . \label{eq:sysXi}
\end{align}
The derived model \eqref{eq:sysXi} links the generated control input sequences with the state of the plant and the buffered control inputs. Due to the stochastic parameter $\rv{\theta}_k$, the augmented system is a Markovian Jump Linear System.

\subsection{Calculation of the Control Law}
\label{sub:sdp}
In the following, we will use the MJLS model \eqref{eq:sysXi} to derive the optimal control law by dynamic programming. 
\begin{remark}
	In literature, there are several solutions to the LQG control problem of MJLS, e.g., \cite{Costa05}. However, these solutions cannot be applied directly since the mode, i.e., $\rv{\theta}_k$ is supposed to be known at time step $k$. In our problem $\rv{\theta}_k$ is only known with a delay of one time step. Another difference is that the weighting matrices $\widetilde{\mat{Q}}_k$ and $\widetilde{\mat{R}}_k$ are stochastic.
\end{remark}
We define the minimal expected cost-to-go $J^*_k$ by
\begin{align}
  J^*_k &= \underset{\vec{U}_k}{\rm min} \E \left\{ C_k + J^*_{k+1} | \mathcal{I}_k \right\} \ , \label{eq:optimality} \\
	J^*_K &= \E \left\{ C_K | \mathcal{I}_K \right\} \ . \label{eq:optimalityBegin}
\end{align}
According to dynamic programming theory \cite{Bertsekas00} it holds, that
\begin{align}
	J^{*}_0 = \underset{\vec{U}_{0:K-1}}{\rm{min}} C^K_0 \ , \no
\end{align}
where $C^K_0$ denotes the expected cumulated cost \eqref{eq:costFunction}.
%
Before we can use \eqref{eq:optimality}, we have to express the stage cost (\ref{eq:stageCostN}) - (\ref{eq:stageCostk}) in terms of the augmented system state $\rvec{\xi}_k$. It holds, that
\begin{align}
	C_K &= \rvec{x}_K\tr \mat{Q}_K \rvec{x}_K = \rvec{\xi}_K\tr \begin{bmatrix} \mat{Q}_K & \mat{0} \\ \mat{0} & \mat{0} \end{bmatrix} \rvec{\xi}_K  = \rvec{\xi}_K\tr \widetilde{\mat{Q}}_K \rvec{\xi}_K, \\
	C_k &= \rvec{x}_k\tr \mat{Q}_k \rvec{x}_k + \vec{u}_k\tr \mat{R}_k \vec{u}_k \no \\
			&= \rvec{x}_k\tr \mat{Q}_k \rvec{x}_k 
			     + \left(\mat{H}_{k} \vec{\eta}_k + \mat{J}_{k} \vec{U}_k\right)\tr \mat{R}_k \left(\mat{H}_{k} \vec{\eta}_k + \mat{J}_{k} \vec{U}_k\right) \no \\
		  &= \begin{pmatrix}
		  			\rvec{x}_k \\ \vec{\eta}_k
		  		\end{pmatrix}\tr
		  		\begin{bmatrix}
		  			\mat{Q}_k & \mat{0} \\
		  			\mat{0}   & \mat{H}_{k}\tr \mat{R}_k \mat{H}_{k}
		  		\end{bmatrix}
  	  		\begin{pmatrix}
		  			\rvec{x}_k \\ \vec{\eta}_k
		  		\end{pmatrix} 
		  		 + \vec{U}_k\tr \mat{J}_{k}\tr \mat{R}_k  \mat{J}_{k} \vec{U}_k \no \\
		  &= \rvec{\xi}_k\tr {\widetilde{\mat{Q}}}_{k} \rvec{\xi}_k + \vec{U}_k\tr {\widetilde{\mat{R}}}_{k} \vec{U}_k  \no \ ,
\end{align}
where we used that $\E \left\{ \mat{H}_k\tr \mat{R}_k \cdot \mat{J}_k \right\} = 0$ and introduced the definitions
\begin{align*}
	\widetilde{\mat{Q}}_K = &
		\begin{bmatrix} 
			\mat{Q}_K & \mat{0} \\ \mat{0} & \mat{0} 
		\end{bmatrix} \no \ , \ \ \ \ \ 
	\widetilde{\mat{Q}}_{k} = 
		\begin{bmatrix}
			\mat{Q}_k & \mat{0} \\
			\mat{0}   & \mat{H}_{k}\tr \mat{R}_k \mat{H}_{k}
		\end{bmatrix}\ , \ \ \ \ \
	\widetilde{\mat{R}}_{k} = \mat{J}_{k}\tr \mat{R}_k  \mat{J}_{k} \ .
\end{align*}
%
Starting at time step $K$, the minimal expected cost-to-go are given by \eqref{eq:optimalityBegin}
\begin{align}
	J^*_K 
	&= \E \{ \rvec{\xi}_K\tr \widetilde{\mat{Q}}_{K} \rvec{\xi}_K | \mathcal{I}_K \}  \no
			= \E \left\{ \rvec{\xi}_K\tr \mat{K}_{K} \rvec{\xi}_K | \mathcal{I}_K \right\} \no
\end{align}
\begin{align}
	\text{with} \ \ \ \mat{K}_{K} = \widetilde{\mat{Q}}_{K} \ .
\end{align}
%
With \eqref{eq:optimality}, it holds for the minimal expected cost-to-go at time step $K-1$
\begin{align}
	 J&^*_{K-1} = \min_{\vec{U}_{K-1}} \E \big\{ \rvec{\xi}_{K-1}\tr \widetilde{\mat{Q}}_{K-1} \rvec{\xi}_{K-1} + \vec{U}_{K-1}\tr \widetilde{\mat{R}}_{K-1} \vec{U}_{K-1} 
	   + \ J^*_K | \mathcal{I}_{K-1} \big\} \nonumber \\
	&= \E \left\{ \rvec{\xi}_{K-1}\tr \widetilde{\mat{Q}}_{K-1} \rvec{\xi}_{K-1} | \mathcal{I}_{K-1} \right\} 
   + \min_{\vec{U}_{K-1}} \left[ \E \left\{ \vec{U}_{K-1}\tr \widetilde{\mat{R}}_{K-1} \vec{U}_{K-1} | \mathcal{I}_{K-1} \right\} 
	 + \E \left\{ \rvec{\xi}_K\tr {\mat{K}}_{K} \rvec{\xi}_K | \mathcal{I}_{K-1} \right\} \right] \label{eq:JN-1rec} \\
	 & =  \ \E \big\{ \rvec{\xi}_{K-1}\tr \widetilde{\mat{Q}}_{K-1} \rvec{\xi}_{K-1} | \mathcal{I}_{K-1} \big\} 
	  +  \min_{\vec{U}_{K-1}} \bigg[ \vec{U}_{K-1}\tr \E \big\{ \widetilde{\mat{R}}_{K-1} | \mathcal{I}_{K-1} \big\} \vec{U}_{K-1}  \nonumber \\
	 & \ \ \ +  \E \Big\{ \Big( \widetilde{\mat{A}}_{K-1} \rvec{\xi}_{K-1} + \widetilde{\mat{B}}_{K-1} \vec{U}_{K-1} + \widetilde{\rvec{w}}_{K-1} \Big)\tr \mat{K}_{K} 
    \times  \Big( \widetilde{\mat{A}}_{K-1} \rvec{\xi}_{K-1} + \widetilde{\mat{B}}_{K-1}  \vec{U}_{K-1} + \widetilde{\rvec{w}}_{K-1} \Big)  | \mathcal{I}_{K-1} \Big\} \bigg] \nonumber\\
   & =  \ \E \left\{ \rvec{\xi}_{K-1}\tr \left( \widetilde{\mat{Q}}_{K-1} + \widetilde{\mat{A}}_{K-1}\tr \mat{K}_{K} \widetilde{\mat{A}}_{K-1} \right) \rvec{\xi}_{K-1} | \mathcal{I}_{K-1} \right\} \nonumber \\
    & \ \ \ +  \min_{\vec{U}_{K-1}} \Big[ \vec{U}_{K-1}\tr \E \Big\{  \widetilde{\mat{R}}_{K-1}
    + \widetilde{\mat{B}}_{K-1}\tr \mat{K}_{K} \widetilde{\mat{B}}_{K-1} | \mathcal{I}_{K-1} \Big\}  \vec{U}_{K-1} \nonumber \\
    & \ \ \ +  \ 2 \cdot \E \left\{ \rvec{\xi}_{K-1}\tr | \mathcal{I}_{K-1} \right\} \E \left\{ \widetilde{\mat{A}}_{K-1}\tr \mat{K}_{K} \widetilde{\mat{B}}_{K-1} | \mathcal{I}_{K-1} \right\}  \vec{U}_{K-1} \Big] 
     + \E \left\{ \widetilde{\rvec{w}}_{K-1}\tr \mat{K}_{K} \widetilde{\rvec{w}}_{K-1} | \mathcal{I}_{K-1} \right\} , \label{eq:toDeviate}
\end{align}
where we used that $\E \{ \E \{g(\vec{\xi}_{k+1}) | \mathcal{I}_{k+1} \} |\mathcal{I}_{k} \} = \E \{ g(\vec{\xi}_{k+1}) | \mathcal{I}_{k} \}$ for any function $g(\cdot)$. Furthermore, we used the fact that if $\mathcal{I}_{K-1}$ is given, then $\rvec\xi_{K-1}$ is conditionally independent of $\rv{\theta}_{K-1}$ and, therefore, of $\mat{A}_{K-1}, \mat{B}_{K-1}$, and $\mat{K}_K$. Differentiation of (\ref{eq:toDeviate}) with respect to $\vec{U}_{K-1}$ and setting equal to zero yields
\begin{align}
\vec{U}_{K-1} =  - \left(\E\left\{ \widetilde{\mat{R}}_{K-1} + \widetilde{\mat{B}}_{K-1}\tr \mat{K}_{K} \widetilde{\mat{B}}_{K-1} | \mathcal{I}_{K-1} \right\}\right)^\dagger  
		\times   \E \left\{ \widetilde{\mat{B}}_{K-1}\tr \mat{K}_{K} \widetilde{\mat{A}}_{K-1} | \mathcal{I}_{K-1} \right\}  \E \left\{ \rvec{\xi}_{K-1} | \mathcal{I}_{K-1}\right\} \label{eq:OptU}
\end{align}
\begin{remark}
In \eqref{eq:OptU}, we have used the Moore-Penrose pseudoinverse instead of the regular inverse as the expression $\mat{M} = \E \left\{ \widetilde{\mat{R}}_{K-1} + \widetilde{\mat{B}}_{K-1}\tr \mat{K}_{K} \widetilde{\mat{B}}_{K-1} | \mathcal{I}_{K-1} \right\}$ is in general not regular but positive semidefinite. This results from two facts: 1) if the network has a latency such that a control input sequence cannot arrive before the first $m \in \mathbb{N}_{>0}$ time steps, i.e., $q_0 = \dots = q_{m-1} = 0$ with $q_{i}$ as defined in sec.~\ref{sub:model}, then the first $m-1$ control inputs of each sequence will never be applied and 2) the last $N$ control input sequences $\vec{U}_{K-N:K-1}$ contain control inputs such as $\vec{u}_{K+1|K-1}$ that are supposed to be applied after the terminal time $K$. Therefore, the minimization problem is not well-defined. One way to cope with this problem is to exclude the corresponding control inputs from the system equations. For case 1) this can be easily done by reducing the vector $\vec{\eta}_k$ by the corresponding control inputs and adjusting the system matrices. In case 2) we have to gradually increase the dimension of $\vec{\eta}_k$ and the system matrices from time step $K$ to $K-N$. Then, the system has reached the dimension of the model derived above and the dimensions stay constant. Another way to solve this issue, is to use the Moore-Penrose pseudoinverse, which yields the same results since the kernel of $\mat{M}^{\dagger}\mat{M}$ is equal to the subspace corresponding to the dimensions of the undefined entries of $U_k$. In this paper, we choose the latter description to obtain a concise expression for the optimal controller, which is straightforward to implement. 
\end{remark}
Using (\ref{eq:OptU}) in (\ref{eq:toDeviate}) gives
\begin{eqnarray}
	 J_{K-1}^* &
   =& \E \left\{ \widetilde{\rvec{w}}_{K-1}\tr \mat{K}_{K} \widetilde{\rvec{w}}_{K-1} | \mathcal{I}_{K-1} \right\}  
  + \E \left\{ \rvec{\xi}_{K-1}\tr \mat{K}_{K-1} \rvec{\xi}_{K-1}| \mathcal{I}_{K-1} \right\}  \nonumber \\
  && + \E \left\{ \left( \rvec{\xi}_{K-1}\tr - \E \left\{ \rvec{\xi}_{K-1}\tr | \mathcal{I}_{K-1}\right\} \right) \mat{P}_{K-1} \right. 
   \times \left( \rvec{\xi}_{K-1} - \E \left\{ \rvec{\xi}_{K-1} | \mathcal{I}_{K-1}\right\}\right)| \mathcal{I}_{K-1} \left. \right\} \nonumber \ ,
\end{eqnarray}
with
\begin{align}
	\mat{K}_{K-1} & = \E \left\{ \widetilde{\mat{Q}}_{K-1}+ \widetilde{\mat{A}}_{K-1}\tr \mat{K}_{K} \widetilde{\mat{A}}_{K-1} |\mathcal{I}_{K-1} \right\}- \mat{P}_{K-1} \ , \\
	\mat{P}_{K-1} & = \E \left\{ \widetilde{\mat{A}}_{K-1}\tr {\mat{K}}_{K}\widetilde{\mat{B}}_{K-1} | \mathcal{I}_{K-1} \right\} \nonumber \\
	& \ \ \ \times \left(\E \left\{ \widetilde{\mat{R}}_{K-1} + \widetilde{\mat{B}}_{K-1}\tr {\mat{K}}_{K}\widetilde{\mat{B}}_{K-1} | \mathcal{I}_{K-1} \right\} \right)^\dagger 
		\times \E \left\{ \widetilde{\mat{B}}_{K-1}\tr {\mat{K}}_{K}\widetilde{\mat{A}}_{K-1} | \mathcal{I}_{K-1} \right\}
\end{align}
Considering one more time step, the minimal expected cost-to-go at time step $K-2$ can be calculated by
\begin{align}
	J^*_{K-2} = & \min_{\vec{U}_{K-2}} \left[ \E \left\{ \rvec{\xi}_{K-2}\tr \widetilde{\mat{Q}}_{K-2} \rvec{\xi}_{K-2} + \vec{U}_{K-2}\tr \widetilde{\mat{R}}_{K-2} \vec{U}_{K-2}
	 + J^*_{K-1} | \mathcal{I}_{K-2}, \vec{U}_{K-2} \right\} \right] \\
	= &\E \left\{ \rvec{\xi}_{K-2}\tr \widetilde{\mat{Q}}_{K-2} \rvec{\xi}_{K-2} | \mathcal{I}_{K-2}\right\}  
	 + \min_{\vec{U}_{K-2}} \left[ \vec{U}_{K-2}\tr \E \left\{ \widetilde{\mat{R}}_{K-2} | \mathcal{I}_{K-2}\right \} \vec{U}_{K-2} \right.  \no\\
  & + \left. \E \left\{ \rvec{\xi}_{K-1}\tr \mat{K}_{K-1} \rvec{\xi}_{K-1}| \mathcal{I}_{K-2},\vec{U}_{K-2} \right\} \right] 
	 + \E \left\{ \left( \rvec{\xi}_{K-1}\tr - \E \left\{ \rvec{\xi}_{K-1}\tr | \mathcal{I}_{K-1}\right\} \right) \mat{P}_{K-1} \right. \no \\
  &  \times \left. \left( \rvec{\xi}_{K-1} - \E \left\{ \rvec{\xi}_{K-1} | \mathcal{I}_{K-1}\right\}\right)| \mathcal{I}_{K-2}, \vec{U}_{K-2}\right\} 
   + \E \left\{ \widetilde{\rvec{w}}_{K-1}\tr \mat{K}_{K} \widetilde{\rvec{w}}_{K-1} | \mathcal{I}_{K-2} \right\}  \label{eq:JN-2struct}.
\end{align}
The term with $\mat{P}_{K-1}$ is excluded from the minimization since it is independent of $\vec{U}_{K-2}$ what is justified in Lemma~\ref{lem:ePe} at the end of this section. The structure of \eqref{eq:JN-2struct} and \eqref{eq:JN-1rec} is the same, besides two additional terms that are independent of $\vec{U}_{0:k-1}$. Therefore, minimization over $\vec{U}_{K-2}$ will lead to a $J^*_{K-2}$ of the same structure, so that it follows by an inductive argument that
\begin{align}
  \vec{U}_{k} =&  - \left(\E\left\{ \widetilde{\mat{R}}_{k} + \widetilde{\mat{B}}_{k}\tr \mat{K}_{k+1} \widetilde{\mat{B}}_{k} | \mathcal{I}_{k} \right\}\right)^\dagger 
	 \E \left\{ \widetilde{\mat{B}}_{k}\tr \mat{K}_{k+1} \widetilde{\mat{A}}_{k} | \mathcal{I}_{k} \right\}  \E \left\{ \rvec{\xi}_{k} | \mathcal{I}_{k}\right\}  \label{eq:OptUk}
\end{align}
with
\begin{align}
	\mat{K}_{k} & = \E \left\{ \widetilde{\mat{Q}}_{k}+ \widetilde{\mat{A}}_{k}\tr \mat{K}_{k+1} \widetilde{\mat{A}}_{k} |\mathcal{I}_{k} \right\} 
								- \E \left\{ \widetilde{\mat{A}}_{k}\tr {\mat{K}}_{dk+1}\widetilde{\mat{B}}_{k} | \mathcal{I}_{k} \right\} \nonumber \\
	\times & \left(\E \left\{ \widetilde{\mat{R}}_{k} + \widetilde{\mat{B}}_{k}\tr {\mat{K}}_{k+1}\widetilde{\mat{B}}_{k} | \mathcal{I}_{k} \right\} \right)^\dagger 
	\E \left\{ \widetilde{\mat{B}}_{k}\tr {\mat{K}}_{k+1}\widetilde{\mat{A}}_{k} | \mathcal{I}_{k} \right\} \ . \label{eq:Kk}
\end{align}
With \eqref{eq:OptUk} and \eqref{eq:Kk}, it follows for the minimal expected cost-to-go at time step $k$ that
\begin{align}
	J^*_{k} = &\E \left\{ \rvec{\xi}_{k}\tr {\mat{K}}_{k+1} \rvec{\xi}_{k} | \mathcal{I}_{k} \right\}  
	+ \displaystyle\sum_{i=k}^{K-1}\E \left\{ \widetilde{\rvec{w}}_{i}\tr \mat{K}_{i+1} \widetilde{\rvec{w}}_{i} | \mathcal{I}_{i} \right\}  
	 + \displaystyle\sum_{i=k}^{K-1}\E \left\{ \left( \rvec{\xi}_{i+1}\tr - \E \left\{ \rvec{\xi}_{i+1}\tr | \mathcal{I}_{i+1}\right\} \right) \mat{P}_{i+1} \right. \nonumber \\
  & \hspace{0.35cm} \times \left. \left( \rvec{\xi}_{i+1} - \E \left\{ \rvec{\xi}_{i+1} | \mathcal{I}_{i+1}\right\}\right)| \mathcal{I}_{i} \right\} \label{eq:JkOpt} \ .
\end{align}
The expected values in \eqref{eq:OptUk} and \eqref{eq:Kk} referring to the matrices can be calculated by explicitly conditioning on $\rv{\theta}_{k-1} = j$. This is possible since $\rv{\theta}_{k-1}$ is part of the information set $\mathcal{I}_k$ and, therefore, known at time step $k$. It holds 
\begin{align}
	\vec{U}_{k} & = - \left[ \displaystyle\sum_{i=0}^{N+1} p_{ji} 
				\left(\widetilde{\mat{R}}_{|i} + \widetilde{\mat{B}}_{|i}\tr \E \left\{\mat{K}_{k+1}|\rv{\theta}_{k} = i \right\} \widetilde{\mat{B}}_{|i} \right)\right]^\dagger 
				\cdot \left[ \displaystyle\sum_{i=0}^{N+1} p_{ji}
				\widetilde{\mat{B}}_{|i}\tr \E \left\{ \mat{K}_{k+1}|\rv{\theta}_{k} = i\right\} \widetilde{\mat{A}}_{|i}  \right]  \cdot \E \left\{ \rvec{\xi}_{k} | \mathcal{I}_k \right\} \label{eq:UkSum}\\
				& \overset{\Delta}{=} \mat{L}_k \E \left\{ \rvec{\xi}_{k} | \mathcal{I}_k \right\} \ \label{eq:Lk}
\end{align}
with%
\begin{align}
  \E &\left\{ \mat{K}_k | \rv{\theta}_{k-1} = j \right\} 
  					 = \left[ \displaystyle\sum_{i=0}^{N+1} p_{ji} \left(\widetilde{\mat{Q}}_{|i}+ \widetilde{\mat{A}}_{|i}\tr \E \left\{ \mat{K}_{k+1}| \rv{\theta}_k = i \right\}\widetilde{\mat{A}}_{|i}\right) \right]
						- \left[\displaystyle\sum_{i=0}^{N+1} p_{ji}  \widetilde{\mat{A}}_{|i}\tr \E \left\{ {\mat{K}}_{k+1} | \rv{\theta}_k = i \right\} \widetilde{\mat{B}}_{|i}\right] \nonumber \\
						& \times \left[ \displaystyle\sum_{i=0}^{K+1} p_{ji}  \left( \widetilde{\mat{R}}_{|i} + \widetilde{\mat{B}}_{|i}\tr \E \left\{ {\mat{K}}_{k+1} | \rv{\theta}_k = i \right\} \widetilde{\mat{B}}_{|i}  \right) \right]^\dagger 
						\cdot \left[ \displaystyle\sum_{i=0}^{N+1} p_{ji}  \widetilde{\mat{B}}_{|i}\tr \E \left\{ {\mat{K}}_{k+1} | \rv{\theta}_k = i \right\} \widetilde{\mat{A}}_{|i} \right] \ , \label{eq:KkSum}
\end{align}
where the notation $\mat{X}_{|i}$, with $i \in \mathbb{N}_0$, refers to the matrix $\mat{X}$ (dependent of $\rv{\theta}_k$), where $\rv{\theta}_k$ is set to $i$. The terms $p_{ji}$ indicate the elements of transition matrix $\mat{T}$ from \eqref{eq:P}.

The results derived above can be summarized as follows:
\begin{theorem}
Consider the problem to find an admissible control law with given sequence length $N$ according to \eqref{eq:mu} that minimizes the cost \eqref{eq:costFunction} - \eqref{eq:stageCostk} 
subject to the system dynamics \eqref{eq:sysX}, the measurement equations \eqref{eq:sysY} and \eqref{eq:zk}, and the actuator logic \eqref{eq:uk} - \eqref{eq:ud}. Then,
	\begin{enumerate}
		\item as in standard LQG control, the separation principle holds, i.e., the optimal control law at time step $k$ can be separated into a) an estimator that calculates the conditional expectation $\E \left\{ \rvec{\xi}_k | \mathcal{I}_k \right\}$ and b) into an optimal state feedback controller that utilizes the state feedback matrix $\mat{L}_k$.
		\item the optimal control law is linear in the conditional expectation of the augmented state, i.e., $U_k = \mat{L}_k \E \left\{ \rvec{\xi}_k  | \mathcal{I}_k \right\}$.
		\item the optimal state feedback matrix $\mat{L}_k$ can be calculated by \eqref{eq:UkSum}, whereas the term $\E \left\{ {\mat{K}}_{k+1} | \rv{\theta}_k = i \right\}$ is obtained by the recursion \eqref{eq:KkSum}, which is evolving backwards in time, with initial condition $\E \left\{ \mat{K}_K | \rv{\theta}_{K-1} = i \right\} = \widetilde{\mat{Q}}_K$.
	\end{enumerate}
\end{theorem}

Since $\vec{\eta}_k$ in \eqref{eq:augState} is known at time step $k$, the conditional expectation $\E \left\{ \rvec{\xi}_k | \mathcal{I}_k \right\}$ reduces to $\E \left\{ \rvec{x}_k | \mathcal{I}_k \right\}$, which is equal to the minimum mean squared error estimate of the state $\rvec{x}_k$. In the literature, results on the minimum mean squared error estimator in presence of measurement delays and measurement losses are available, see e.g., \cite{schenato2008optimal} and \cite{moayedi2010adaptive}. It was pointed out in \cite{schenato2008optimal} that the optimal estimator is a time-varying Kalman filter that is extended by a buffer to store old measurements. The buffered measurements are needed to incorporate delayed measurements that are received after measurements with more recent information. This filter is of finite dimension (and, therefore, can be implemented) when the required memory of the buffer is finite. This is the case if every measurement delay is either bounded or infinite\footnote{This formulation is consistent with the one in \cite{schenato2008optimal}, where it is required for the finite optimal filter to exist that a measurement, that is not lost, will arrive within a maximal time.}. In practice, however, it is computational inefficient to process measurements that have been delayed for a very long time and, therefore, the length of the buffer is treated as a design parameter. 

For similar reasons, we assumed in the problem formulation that the length $N$ of the control input sequence is a given design parameter. Otherwise, if the minimization would be carried out over $N$, and we consider networks with data losses, then the resulting length of the control input sequence would be infinite.

\begin{lemma}
\label{lem:ePe}
The term
\begin{align*}
  &\E \left\{ \left( \rvec{\xi}_{k}\tr - \E \left\{ \rvec{\xi}_{k}\tr | \mathcal{I}_{k}\right\} \right) \mat{P}_{k}  
   \times  \left( \rvec{\xi}_{k} - \E \left\{ \rvec{\xi}_{k} | \mathcal{I}_{k}\right\}\right)| \mathcal{I}_{k-1}, \vec{U}_{k-1} \right\} \ ,
\end{align*}
is stochastically independent of the control sequence $\vec{U}_{k-1}$.
\end{lemma}

\begin{Proof}

Consider the system \eqref{eq:sysXi} and the autonomous system, given by
\begin{equation} \hat{\rvec{\xi}}_k = \widetilde{\mat{A}}_{k-1} \hat{\rvec{\xi}}_{k-1} + \widetilde{\rvec{w}}_{k-1} \ ,\label{eq:sysHatXi} \end{equation}
that has same system matrix, initial conditions, noise realizations $\rvec{w}_{0:k-1}$, $\rvec{v}_{0:k-1}$, and network delay realizations $\rv{\tau}^{C\!A}_{{0:k-1}}$, $\rv{\tau}^{SE}_{{0:k-1}}$.
Both systems evolve according to time-variant transformations, which are linear, so that it is possible to find matrices $\hat{\mat{A}}$, $\hat{\mat{B}}$, and $\hat{\mat{C}}$ depending on the realizations of $\rv{\theta}_{0:k-1}$ with
\begin{align}
 \rvec{\xi}_k &= \hat{\mat{A}} \rvec{\xi}_0 + \hat{\mat{B}} \begin{bmatrix} \vec{U}_0\tr, \dots , \vec{U}_{k-1}\tr \end{bmatrix}\tr + \hat{\mat{C}} \begin{bmatrix} \rvec{w}_0\tr, \cdots, \rvec{w}_{k-1}\tr \end{bmatrix}\tr \nonumber \\
\hat{\rvec{\xi}}_k &= \hat{\mat{A}} \rvec{\xi}_0 + \hat{\mat{C}} \begin{bmatrix} \rvec{w}_0\tr, \cdots, \rvec{w}_{k-1}\tr \end{bmatrix}\tr \nonumber \ ,
\end{align}
It holds for the expected values
\begin{align}
\E \left\{ \rvec{\xi}_k | \mathcal{I}_k \right\} &= \hat{\mat{A}} \E \left\{ \rvec{\xi}_0 |\mathcal{I}_k \right\} + \hat{\mat{B}} \begin{bmatrix} \vec{U}_0\tr, \cdots , \vec{U}_{k-1}\tr \end{bmatrix}\tr \nonumber \ , \\ 
\E \left\{ \hat{\rvec{\xi}}_k | \mathcal{I}_k \right\} &= \hat{\mat{A}} \E \left\{ \rvec{\xi}_0 |\mathcal{I}_k \right\} \nonumber \ ,
\end{align}
where $\hat{\mat{A}}$ and $\hat{\mat{B}}$ are known since the information vector $\mathcal{I}_k$ includes $\rv{\theta}_{0:k-1}$. Defining the estimation errors $\rv{e}_k = \rvec{\xi}_k - \E \left\{ \rvec{\xi}_k | \mathcal{I}_k \right\}$ and $\hat{\rv{e}}_k = \hat{\rvec{\xi}}_k - \E \left\{ \hat{\rvec{\xi}}_k | \mathcal{I}_k \right\}$, it holds that
\begin{align}
\rv{e}_k  = \hat{\mat{A}} \left( \rvec{\xi}_0 - \E \left\{ \rvec{\xi}_0 |\mathcal{I}_k \right\} \right) + \hat{\mat{C}} \begin{pmatrix} \rvec{w}_0\tr & \cdots & \rvec{w}_{k-1}\tr \end{pmatrix}\tr \ , \nonumber \\
 \hat{\rv{e}}_k = \hat{\mat{A}} \left( \rvec{\xi}_0 - \E \left\{ \rvec{\xi}_0 |\mathcal{I}_k \right\} \right) + \hat{\mat{C}} \begin{pmatrix} \rvec{w}_0\tr & \cdots & \rvec{w}_{k-1}\tr \end{pmatrix}\tr \ . \nonumber
\end{align}
Therefore, the errors are identical and since
\begin{align}
	\rv{e}_k = \hat{\rv{e}}_k = \hat{\rvec{\xi}}_k - \E \left\{ \hat{\rvec{\xi}}_k | \mathcal{I}_k \right\} = \hat{\rvec{\xi}}_k - \E \left\{ \hat{\rvec{\xi}}_k | {z}_{0:k}, \rv{\theta}_{0:k-1} \right\} \ , \nonumber
\end{align}
it holds that the error $\rv{e}_k$ is independent of $\vec{U}_{k-1}$ and that 
\begin{align}
	\E \left\{ \rv{e}_k\tr \rv{e}_k | \mathcal{I}_{k-1}, \vec{U}_{k-1} \right\} = \E \left\{ \rv{e}_k\tr \rv{e}_k | z_{0:k-1}, \rv{\theta}_{0:k-2} \right\} \ .
\end{align}
If $\rv{\theta}_{k-2}$ is given, then $\rv{\theta}_{k-1}$ and $\rv{\theta}_{k}$ are conditionally independent of $z_{k-1}$, $\vec{U}_{k-2}$ and $\vec{U}_{k-1}$. It follows that
\begin{align}
\E \left\{\mat{P}_k | \mathcal{I}_{k-1}, \vec{U}_{k-1} \right\} = \E \left\{\mat{P}_k | \rv{\theta}_{k-2} \right\}
\end{align}
and finally
\begin{align}
	\E \left\{ \rv{e}_k\tr \mat{P}_k \rv{e}_k | \mathcal{I}_{k-1}, \vec{U}_{k-1} \right\} = \E \left\{ \rv{e}_k\tr \mat{P}_k \rv{e}_k  | z_{0:k}, \rv{\theta}_{0:k-1} \right\} \  \nonumber
\end{align}
concludes the proof.
\end{Proof}

\begin{figure*}[h]
	\centering
		\includegraphics[width=0.6\textwidth]{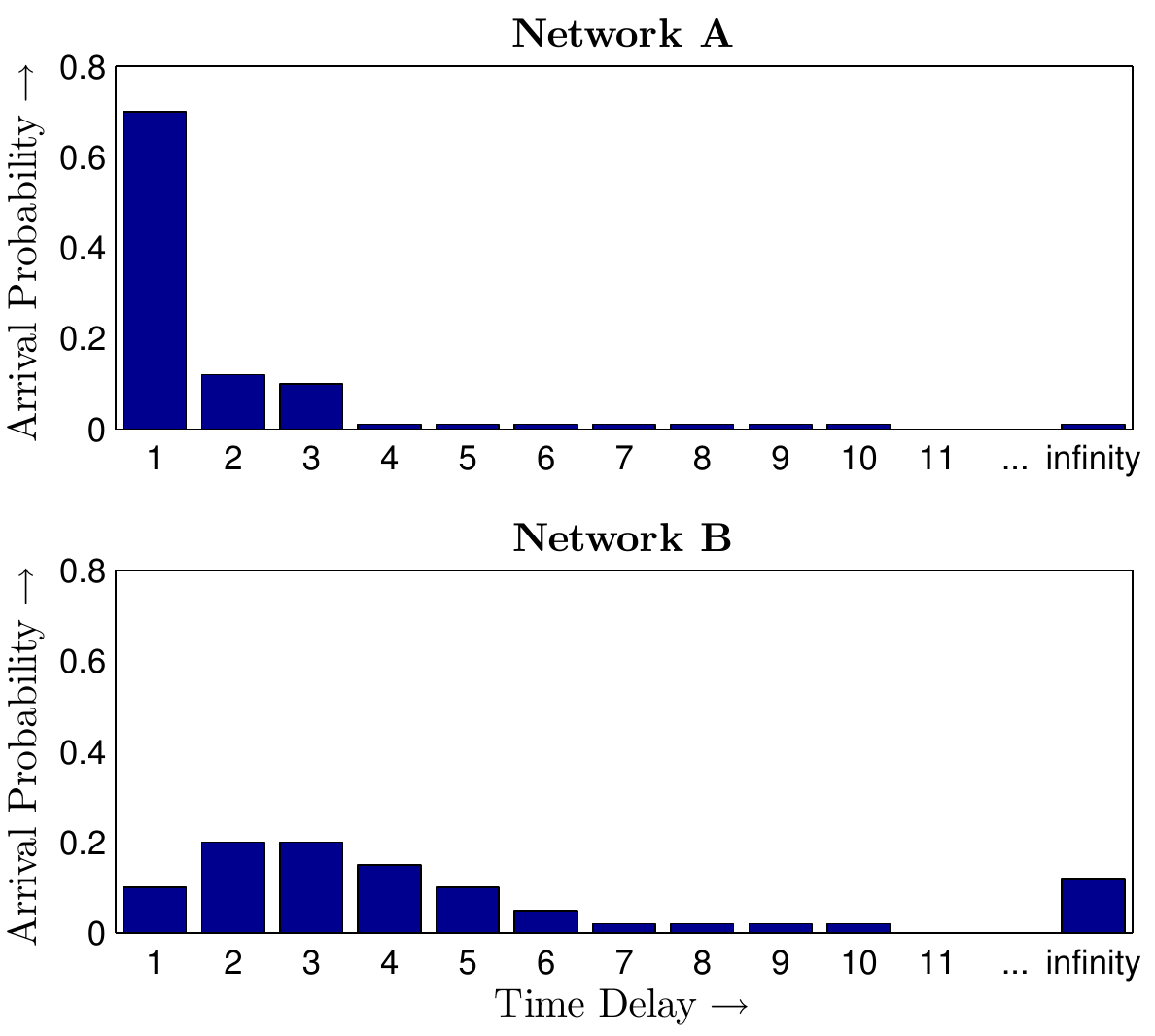}
	\caption{Probability density functions of the arrival probability of data packets over time  delays.}
	\label{fig:Networks}
\end{figure*}

%% file: 50_simulations.tex
In this section, we compare the performance of the proposed optimal controller with the approaches presented in \cite{gupta2006receding} and \cite{moayedi2011lqg} by means of simulations with a double integrator. For simulation, the system parameters of \eqref{eq:sysX} and \eqref{eq:sysY} are chosen with
\begin{align*}
	&\mat{A} = \begin{bmatrix} 1 & 1 \\ 0 & 1 \end{bmatrix}, 
	\ \ \ \ \mat{B} = \begin{bmatrix} 0 \\ 1 \end{bmatrix}, 
	\ \ \ \ \mat{C} = \begin{bmatrix} 1 & 0\end{bmatrix} ,
\end{align*}
and the weighting matrices of the cost function \eqref{eq:costFunction}, the initial condition and the noise covariances are set to
\begin{align*}
	& \mat{Q} = \begin{bmatrix} 1 & 0 \\ 0 & 1 \end{bmatrix},  \ \ \ \mat{R} = 1 \ ,	\bar{x}_0 = \begin{bmatrix} 100 \\ 0 \end{bmatrix} , \ \ \ \bar{\mat{P}}_0 = \begin{bmatrix} 0.5^2 & 0 \\ 0 & 0.5^2 \end{bmatrix} \ ,\\
	& \E\left\{\rvec{w}_k\tr \rvec{w}_k\right\} = \begin{bmatrix} 0.1^2 & 0 \\ 0 & 0.1^2 \end{bmatrix}, \ \ \ \E\left\{\rv{v}_k\tr \rv{v}_k\right\} = 0.2^2 \ .
\end{align*}
In the simulation, we use two different models of the network connections. The probability density functions of the delay distributions of both networks are depicted in fig.~\ref{fig:Networks}. Thereby, Network A has a better transmission quality than Network B as, first, the probability of a small time delay is significantly higher and, second, the loss probability, i.e. the probability of an infinite delay, is much smaller.
\begin{figure*}[h]
	\centering
		\includegraphics[width=0.6\textwidth]{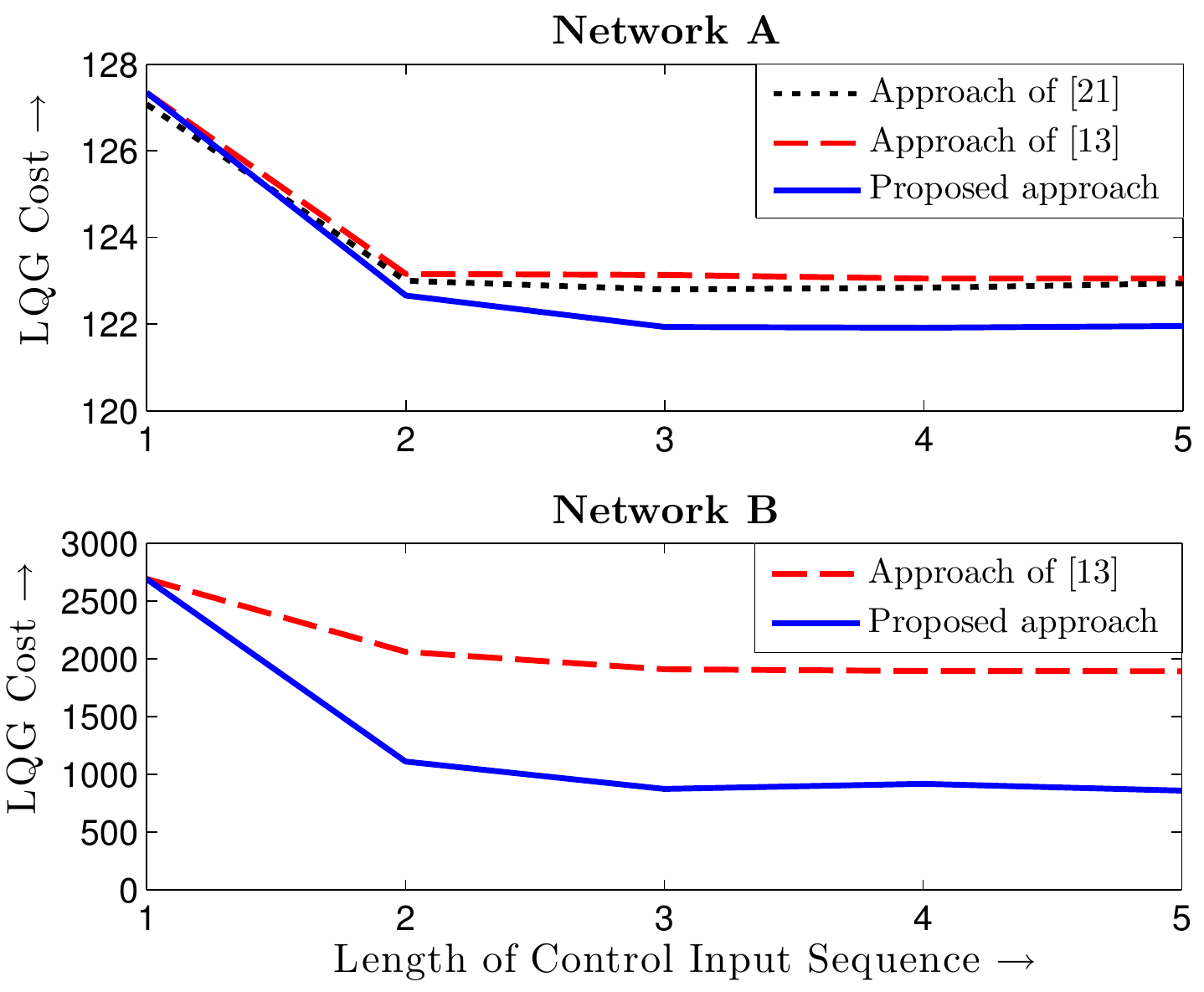}
	\caption{Resulting averaged cumulated LQG cost for Network A and Network B for different lengths of the control input sequences.}
	\label{fig:Cost}
\end{figure*}
We assume that the probability density function of the controller-actuator network is the same as the one of the sensor-controller network. 
To obtain the minimum mean squared error estimate of the state, i.e., the conditional expectation in \eqref{eq:Lk}, we employed the filter described in \cite{moayedi2010adaptive}. The filter is chosen so that it can process measurements with a delay of more than 10 time steps and, therefore, yields the optimal state estimate. If the buffer of the actuator runs empty the actuator applies the default control input \mbox{$\vec{u}^d$ = 0}.

For each controller and network and for different length $N$ of the control input sequence, we conduct 500 Monte Carlo simulation runs over 40 time steps and calculate the average of the cumulated cost \eqref{eq:costFunction}. The results are shown in fig.~\ref{fig:Cost}.
The proposed controller leads to the lowest cost for both networks. For Network A, the difference between the three controllers is only small. This results from the good transmission quality of the network, in particular, from the high probability that a data packet will arrive at the actuator without any delay. This is extremely beneficial for the approach in \cite{gupta2006receding}, where only undelayed packets can be used by the actuator, and for the approach in \cite{moayedi2011lqg}, which is based on the approximation that the controller uses only the state information \textit{deemed available to the actuator} (see \cite{moayedi2011lqg} for more details). 

For Network B the approach of \cite{moayedi2011lqg} is not depicted as it is not able to stabilize the system, i.e., leads to cost that are several magnitudes higher than the cost of the other controllers. Compared to Network A the cumulated cost of the proposed controller and the controller from \cite{gupta2006receding} are higher, what results from the worse transmission quality, which unavoidably degrades the performance. However, for a sequence length of $N>1$ the proposed approach leads to half of the cost compared to the approach of \cite{gupta2006receding}.

%% file: 60_conclusions.tex
We presented an optimal solution to the sequence-based LQG control problem for NCS with TCP-like network connections. In contrast to former work, we were able to optimally consider time-varying packet delays in the sequence-based controller design.

Future work will be concerned with a derivation of stability conditions for the proposed controller. It seems reasonable to assume that the stability region is larger than the one of the approaches in \cite{gupta2006receding} and \cite{moayedi2011lqg}, but this has still to be proved. Furthermore, we seek to include a simple cost model of the network into the cumulated cost function so that the controller also considers the power consumption of a transmission. This way, the minimization is also carried out over the length of the control input sequences and the controller may, e.g., decide not to send a sequence since the buffered sequence is still good enough. Finally, it is an aim to relax the assumptions of the TCP-like network connection in favor of realistic TCP connections with possibly delayed acknowledgements.